\documentclass[twocolumn,twoside]{IEEEtran}

\usepackage{mathpazo}
\usepackage{times}
\usepackage{amsmath}
\usepackage{amsfonts}
\usepackage{latexsym}
\usepackage{amssymb}
\usepackage{cite}

\usepackage{upref}
\usepackage{theorem}
\usepackage{psfrag}
\usepackage{color}

\usepackage[pdftex]{graphicx}

\newtheorem{alg}{Algorithm}

\newtheorem{theorem}{Theorem}
\newtheorem{lem}{Lemma}
\newtheorem{clm}[lem]{Claim}
\newtheorem{definition}{Definition}

\newtheorem{coro}{Corollary}

\newtheorem{conjecture}{Conjecture}
\newtheorem{proposition}{Proposition}

\newcommand{\BA}{\begin{alg}} \newcommand{\EA}{\end{alg}}
\newcommand{\BE}{\begin{enumerate}} \newcommand{\EE}{\end{enumerate}}
\newcommand{\BT}{\begin{theorem}} \newcommand{\ET}{\end{theorem}}
\newcommand{\BL}{\begin{lem}} \newcommand{\EL}{\end{lem}}
\newcommand{\BCM}{\begin{clm}} \newcommand{\ECM}{\end{clm}}
\newcommand{\BCR}{\begin{coro}} \newcommand{\ECR}{\end{coro}}
\newcommand{\BP}{\begin{proposition}} \newcommand{\EP}{\end{proposition}}

\newcommand{\BI}{\begin{itemize}} \newcommand{\EI}{\end{itemize}}

\def\FullBox{\hbox{\vrule width 8pt height 8pt depth 0pt}}
\newcommand{\qed}{\;\;\;\FullBox}
\newenvironment{prf}{\noindent{\bf Proof:~~}}{\(\qed\)}
\newcommand{\BPF}{\begin{prf}} \newcommand {\EPF}{\end{prf}}
\newenvironment{proofof}[1]{\noindent{\bf Proof of {#1}.~}}{\endprf}
\newcommand{\BPFOF}{\begin{proofof}} \newcommand {\EPFOF}{\end{proofof}}

\newcommand{\BEQN}{\begin{eqnarray}}\newcommand{\EEQN}{\end{eqnarray}}
\newcommand{\BEQ}{\begin{equation}} \newcommand{\EEQ}{\end{equation}}

\newcommand{\eat}[1]{}
\newcommand{\eps}{\varepsilon}

\begin{document}

\title{Minimizing the alphabet size of erasure codes with restricted decoding sets}
\author{\large
Mira Gonen,
Ishay Haviv,
Michael Langberg,
Alex Sprintson
\thanks{Mira Gonen is with the Department of Computer Science, Ariel University, Ariel 40700, Israel (e-mail: mirag@ariel.ac.il).}
\thanks{Ishay Haviv is with the School of Computer Science, The Academic College of Tel Aviv-Yaffo, Tel Aviv 61083, Israel.}
\thanks{Michael Langberg is with the Department
 of Electrical Engineering, State University of New-York at Buffalo, Buffalo, NY 14260, USA (e-mail: mikel@buffalo.edu). Work supported in part by NSF grant 1909451.}
\thanks{Alex Sprintson is with the Department of Electrical and Computer Engineering, Texas A\&M University, College Station, TX 77843-3128, USA (e-mail:  spalex@tamu.edu). Work supported in part by NSF grants 1642983 and 1718658.}
  }



\maketitle

\addtolength{\textfloatsep}{-1.01\baselineskip}
\addtolength{\dbltextfloatsep}{-1.01\baselineskip}
\addtolength{\belowdisplayskip}{-1.01ex}
 \addtolength{\abovedisplayshortskip}{-1.01ex}
 \addtolength{\belowdisplayshortskip}{-1.01ex}

\begin{abstract}
A Maximum Distance Separable code over an alphabet $F$ is defined via an encoding function $C:F^k \rightarrow F^n$ that allows to retrieve a message $m \in F^k$ from the codeword $C(m)$ even after erasing any $n-k$ of its symbols. The minimum possible alphabet size of general (non-linear) MDS codes for given parameters $n$ and $k$ is unknown and forms one of the central open problems in coding theory. The  paper initiates the study of the alphabet size of codes in a {\em generalized} setting where the coding scheme is required to handle a pre-specified subset of all possible erasure patterns, naturally represented by an $n$-vertex $k$-uniform hypergraph.
We relate the minimum possible alphabet size of such codes to the strong chromatic number of the hypergraph and analyze the tightness of the obtained bounds for both the linear and non-linear settings.
We further consider variations of the problem which allow a small probability of decoding error.
\end{abstract}

\section{Introduction}



Maximum Distance Separable codes are known to play an important and influential role in the area of coding theory.
An MDS code over an alphabet $F$ is defined via an encoding function $C:F^k \rightarrow F^n$ that allows to retrieve a message \mbox{$m \in F^k$} from the codeword $C(m)$ even after erasing any $n-k$ of its symbols. Equivalently, the Hamming distance between any two distinct codewords is at least $n-k+1$.
The well-known Singleton bound implies that MDS codes are optimal with respect to the number of erasures that they can handle.
However, the minimum possible alphabet size of such codes for given parameters $n$ and $k$ is unknown and forms a central open question in coding theory (see Conjectures~\ref{con:MDS_new} and~\ref{con:MDS_linear_new}).

The present paper initiates the study of the alphabet size of codes in a {\em generalized} setting where the coding scheme is required to handle a pre-specified subset of all possible erasure patterns.
Such scenarios arise naturally in the distributed storage settings in which there is a need to rebuild the entire data set by contacting one of the pre-specified sets of storage nodes, referred to as a \emph{recovery group}. The desired set of recovery groups is determined based on the network configuration, the reliability of storage nodes, as well as the network access patterns. Similar constraints appear in \emph{availability codes} \cite{8007111}, which have recently attracted interest from the research community. However, in availability codes each repair group is used for retrieving a single symbol, whereas we are focusing on retrieving the entire set of $k$ symbols.

The set of erasure patterns is naturally represented by an $n$-vertex $k$-uniform hypergraph in which the vertices represent the $n$ coordinates of the codewords and the (hyper)edges\footnote{For clarify, in the rest of the paper we refer to the hyperedges of a hypergraph as \emph{edges}.} correspond to the possible sets of locations of unerased symbols (i.e., {\em decoding sets}). For a given uniform hypergraph $G$ we are interested in minimizing the size of the alphabet over which there exists a coding scheme with respect to the erasure patterns defined by $G$.

\begin{definition}[The $q$ parameter]\label{def:q_parameter}
Let $G=([n],E)$ be a $k$-uniform hypergraph on the vertex set $[n]=\{1,\ldots,n\}$.
Let $q(G)$ denote the smallest size $q$ of an alphabet $F$ for which there exist an encoding function
\[C: F^k \rightarrow F^n\]
and a decoding function
\[ D: (F \cup \{\perp\})^n \rightarrow F^k \]
such that for every edge $e \in E$ and every message $m \in F^k$ it holds that
\[ D(C_e(m)) = m.\]
Here, $C_e(m)$ stands for the word obtained from the codeword $C(m)$ by replacing the symbols in the locations of $[n] \setminus e$ by the erasure symbol $\perp$.

Similarly, let $q_{lin}(G)$ denote the smallest prime power $q$ for which there exist {\em linear} encoding and decoding functions as above when $F$ is a field of size $q$.
\end{definition}

Observe that for the complete $n$-vertex $k$-uniform hypergraph, denoted by $\kappa_{n,k}$, the values of $q(\kappa_{n,k})$ and $q_{lin}(\kappa_{n,k})$  are equal to the minimum alphabet sizes of general and linear $(n,k)$ MDS codes, respectively. We state below the MDS conjectures for general and for linear codes (see, e.g.,~\cite{B52,S55,MDS-linear,Huntemann_thesis}).

\begin{conjecture}[MDS Conjecture for general codes]
\label{con:MDS_new}
For given integers $k < q \neq 6$, let $n(q,k)$ be the largest integer $n$ such that $q(\kappa_{n,k}) \leq q$. Then,
\begin{equation}\label{eq:MDS}
n(q,k) \le\left\{
	\begin{array}{ll}
		q+2  & \mbox{if } 4|q \mbox{ and } k\in \{3, q-1\}\\
		q+1 & \mbox{otherwise.}
	\end{array}
\right.
\end{equation}
\end{conjecture}

\begin{conjecture}[MDS Conjecture for linear codes]
\label{con:MDS_linear_new}
For given integers $k < q$ where $q$ is a prime power, let $n(q,k)$ be the largest integer $n$ such that $q_{lin}(\kappa_{n,k}) \leq q$. Then,
\begin{equation}\label{eq:MDS}
n(q,k) \le\left\{
	\begin{array}{ll}
		q+2  & \mbox{if } q \mbox{ is even and } k\in \{3, q-1\}\\
		q+1 & \mbox{otherwise.}
	\end{array}
\right.
\end{equation}
\end{conjecture}

%

\noindent
Note that in Conjecture~\ref{con:MDS_linear_new} the right-hand side is known to form a lower bound on the left-hand side, and that the special case of the above conjectures for $k=2$ is known to hold (see, e.g.,~\cite{Huntemann_thesis}). The MDS Conjecture for linear codes over prime fields has been proven by S. Ball in his seminal paper  \cite{ball2012sets}.

In this work we aim to study the behavior of the $q$ parameter for general sub-hypergraphs of $\kappa_{n,k}$.
Our results imply strong relations between the $q$ parameter of uniform hypergraphs and their chromatic number.
A {\em valid} coloring of a hypergraph $G$ is an assignment of colors to its vertices so that the vertices of each edge are assigned to distinct colors.
This is at times referred to as a {\em strong} coloring and is consistent with the notion of graph coloring (i.e., the coloring of hypergraphs with edges of size two).
The chromatic number $\chi(G)$ of $G$ is the minimum number of colors that allows a valid coloring of $G$.

\section{Our results}

In what follows we give an overview of our results. The proofs are addressed in Sections~\ref{sec:proofs_q} and \ref{sec:q_eps}.

\vspace{2mm}

\noindent
\subsection{ Relation between $q$ and $\chi$}

We start with the following upper bounds.

\BT\label{the:reduction_new}
For every $k$-uniform hypergraph $G$,
\[q(G) \leq q(\kappa_{\chi(G),k}) \mbox{ ~~and~~  } q_{lin}(G) \leq q_{lin}(\kappa_{\chi(G),k}).\]
In particular,  \[q(G) \leq q_{lin}(G) \leq [\chi(G)-1]_{pp}.\]
Here, for an integer $x$, $[x]_{pp}$ represents the smallest prime power that is greater or equal to $x$.
\ET

Theorem~\ref{the:reduction_new} formalizes the natural intuition that for {\em simple} collections of erasure patterns $G$, i.e., the setting in which  $\chi(G)$ is small, a {\em small} alphabet size $q$ suffices for a suitable erasure code. The question we next consider is whether the upper bound provided by Theorem~\ref{the:reduction_new} is tight.

\vspace{2mm}

\noindent
\subsection{ Tightness of Theorem~\ref{the:reduction_new}, the case $k \geq 3$}
The following result shows that Theorem~\ref{the:reduction_new} is not tight in general.

\BP
\label{prop:Fano}
There exists a $3$-uniform hypergraph $G$ with $q_{lin}(G) = q(G) = 2$ and yet $q(\kappa_{\chi(G),3}) \geq 5$.
\EP

We further show that for every $k \geq 3$ the chromatic number of $k$-uniform hypergraphs can be significantly larger than their $q$ parameter (even while restricted to the linear setting).
This implies a large gap between the $q$ parameter and its upper bound provided by Theorem~\ref{the:reduction_new}.

\BP
\label{prop:GL_new}
For every $k\ge 3$ and every prime power $q$, there exists a $k$-uniform hypergraph $G$ with $q_{lin}(G) \leq q$ and yet $\chi(G) \geq \frac{q^k-1}{q-1}$.
\EP

For $k=2$, the question at hand seems to be more challenging.
Here we ask whether there exists a graph $G$ for which $q(G)$ is significantly smaller than $q_{lin}(G)$ which in turn is known to be at most $[\chi(G)-1]_{pp}$.
We thus study the relationship between $q(G)$ and $\chi(G)$.
Our results for this case are outlined below.

\vspace{2mm}

\noindent
\subsection{ Tightness of Theorem~\ref{the:reduction_new}, the case $k = 2$}
To study the relationship between $q$ and $\chi$ for the case $k=2$, we define the following graph family.

\begin{definition}[The graph family $G_{q}$]
\label{def:Gq}
For an integer $q$, let $G_q$ be the graph whose vertex set consists of all the balanced vectors of length $q^2$ over $[q]$, that is, the vectors $u \in [q]^{q^2}$ such that $|\{i \in [q^2] \mid u_i =j\}|=q$ for every $j \in [q]$, where two vertices $u=(u_1,\dots,u_{q^2})$ and $v=(v_1,\dots,v_{q^2})$ are adjacent if the collection of pairs $\{(u_i,v_i)\}_{i \in [q^2]}$ is equal to $[q]\times [q]$.
\end{definition}

The graph family $G_q$ is extremal with respect to the $q$ parameter in the sense given by the following lemma.

\BL
\label{lem:subgraph_new}
For every integer $q$,
\begin{enumerate}
  \item $q(G_q) \leq q$ and
  \item $\chi(G) \leq \chi(G_q)$ for every graph $G$ with $q(G)=q$.
\end{enumerate}
\EL

By Lemma~\ref{lem:subgraph_new}, the challenge of obtaining graphs with chromatic number much larger than the $q$ parameter reduces to the study of the chromatic number $\chi(G_q)$ of $G_q$.
We first show that if we require the coding scheme of $G_q$ to be linear then the size of the used alphabet cannot be smaller than $[\chi(G_q)-1]_{pp}$ (implying that Theorem~\ref{the:reduction_new} is tight in this case). Recall that this is in contrast to the situation of $k \geq 3$ (see Proposition~\ref{prop:GL_new}).

\BP
\label{prop:linear:2_new}
For every integer $q$,
\[q_{lin}(G_q) = [\chi(G_q)-1]_{pp}.\]
\EP

And what about non-linear codes? Is $q(G_q)$ significantly smaller than $\chi(G_q)$?
For $q=2$, it is not difficult to see that $\chi(G_q)=3=q(G_q)+1$.
However, an exhaustive analysis due to~\cite{LL05} shows that for $q=3$ it holds that $\chi(G_q) =6 > q(G_q)+1$.
For general values of $q$, we provide the bounds stated below.
We use here the notation $n(q,2)$ which stands for the largest integer $n$ for which there exists an MDS code of length $n$ and dimension $2$ over an alphabet of size $q$.
Note that if $q$ is a prime power then $n(q,2) = q+1$.

\BP
\label{prop:2:nonlinear_new}
For every integer $q$,
\[n(q,2) \leq \chi(G_{q}) \leq {q+1\choose 2}.\]
In particular, if $q$ is a prime power then
\[q+1 \leq \chi(G_{q}) \leq {q+1\choose 2}.\]
\EP

It is interesting to understand the asymptotic behavior of $\chi(G_q)$ as a function of $q$.
In an attempt to shed some light on this question, we provide a couple of related results, described next.

\vspace{2mm}

\noindent
\subsection{ Understanding $\chi(G_q)$}


Firstly, we consider a natural family of independent sets of $G_q$ which we refer to as {\em canonical} independent sets.
The canonical independent set $A_{i,j}$ associated with two distinct indices $i,j \in [q^2]$ is the set of all vertices $u \in [q]^{q^2}$ of $G_q$ that satisfy $u_i = u_j$. For a prime power $q$, it can be seen that $A_{i,j}$ is an independent set of maximum size in $G_q$. In fact, this type of independent set is used to obtain the upper bound in Proposition~\ref{prop:2:nonlinear_new}. However, we show that if we restrict ourselves to colorings of $G_q$ whose color classes are all contained in canonical independent sets then the number of used colors has to be quadratic in $q$ (see Proposition~\ref{prop:canonical}) and is thus close to the upper bound of Proposition~\ref{prop:2:nonlinear_new}.

Secondly, we focus on the subgraph of $G_q$ induced by the vertices whose vectors in $[q]^{q^2}$ form concatenations of $q$ permutations of $[q]$. It is shown that the chromatic number of this subgraph is only linear in $q$, corresponding now to the lower bound of Proposition~\ref{prop:2:nonlinear_new}. Intuitively speaking, this might hint that the difficulty in coloring the graph $G_q$ using few colors comes from the `less-structured areas' of the graph (see Proposition~\ref{prop:subgraphs_G_q}).

We conclude our work with a further extension of the $q$ parameter to an error-tolerant model.

\vspace{2mm}

\noindent
\subsection{ Allowing an error in decoding}
In this last study, for a given uniform hypergraph $G$ and an error parameter $\eps > 0$, we are interested in minimizing the size of the alphabet over which there exists a coding scheme with respect to the erasure patterns defined by $G$ that guarantees success probability at least $1-\eps$. We first study the setting in which the success probability is taken over the uniform distribution on the messages. This is given formally in the following definition.


\begin{definition}[The $q_\eps$ parameter]\label{def:q_eps_parameter}
Let $G=([n],E)$ be a $k$-uniform hypergraph on the vertex set $[n]$ and let $\eps >0$.
Let $q_\eps(G)$ denote the smallest size $q$ of an alphabet $F$ for which there exist an encoding function
$C: F^k \rightarrow F^n$
and a decoding function
$D: (F \cup \{\perp\})^n \rightarrow F^k$
such that for every edge $e \in E$ it holds that
\[\Pr_{m}[D(C_e(m))=m] \geq 1-\eps,\]
where $m$ is uniformly chosen from $F^k$.
\end{definition}
\noindent
Notice that the $q$ parameter given in Definition~\ref{def:q_parameter} coincides with Definition~\ref{def:q_eps_parameter} when $\eps = 0$.
Similar to Theorem~\ref{the:reduction_new}, the following holds.

\BT\label{the:reduction_new_eps}
For every $\eps \geq 0$ and a $k$-uniform hypergraph $G$,
$$q_\eps(G) \leq q_\eps(\kappa_{\chi(G),k}) \leq q(\kappa_{\chi(G),k}).$$
\ET

We focus our study on the case $k=2$. To demonstrate Definition~\ref{def:q_eps_parameter}, we consider the complete graph $\kappa_{n,2}$.
For $\eps=0$, this takes us to $2$-dimensional MDS codes.
However, if the probability of success is slightly relaxed it turns out that the required alphabet size can be reduced.
For example, in Proposition~\ref{p:q3} of Section~\ref{sec:q_eps} we show that
\begin{itemize}
\item for $\eps = \frac{1}{3}$, $q_{\eps}(\kappa_{20,2}) \leq 3$ whereas $q(\kappa_{20,2}) =19$, \vspace{1mm}
\item for $\eps = \frac{1}{4}$, $q_{\eps}(\kappa_{7,2}) \leq 4$ whereas $q(\kappa_{7,2}) =6$, and \vspace{1mm}
\item for $\eps = \frac{1}{6}$, $q_{\eps}(\kappa_{6,2}) \leq 6$ whereas $q(\kappa_{6,2}) = 7$.
\end{itemize}
Notice that, for $k=2$, $\eps >0$ implies $\eps \geq \frac{1}{q^2}$ (as our error is measured over a sample space of size $q^2$). Thus, as a first step in understanding $q_{\eps}$, in the analysis above and those that follow, we study $\eps$ for the intermediate value of $\frac{1}{q}$. A full study addressing $q_{\eps}$ for general $\eps$ is left for future work.

In Section~\ref{sec:q_eps}, we study the $q_\eps$ parameter  for general graphs $G$ (with $k=2$ and $\eps=1/q$).
As in our study of the $q$ parameter, we employ the tool of universal graphs (i.e., an analog to the graph family of Definition~\ref{def:Gq}) and apply it for the error-tolerant setting. See Propositions~\ref{p:Gqe},~\ref{p:Hqe_cyclic}, and~\ref{p:Hqe} in Section~\ref{sec:q_eps}.

Finally, in Section~\ref{sec:average}, we further extend the notion of error to allow an average error $\tilde\eps$ over both the messages and the edges in $E$. Here, the decoding error is computed assuming a uniform set of $k$ messages and a uniform decoding edge in $E$ (see Definition~\ref{def:q_avr_eps_parameter} in Section~\ref{sec:average}). Roughly speaking, we show in Theorem~\ref{the:average} that this last notion of error allows significant flexibility in the sense that for a given $\tilde\eps$ the relaxed $q_{\tilde\eps}$ parameter for any sized clique is bounded by approximately $1/\tilde\eps$.

\begin{figure}[b]
    \begin{center}
            \resizebox{80mm}{!}{\includegraphics[scale=0.25, viewport=0 14 1000 527]{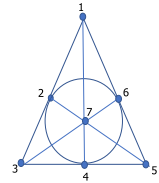}}
    \end{center}
    \caption{Illustration of the Fano matroid.}\label{fig:Fano_new}
\end{figure}

\section{The $q$ Parameter}
\label{sec:proofs_q}

\subsection{Proof of Theorem~\ref{the:reduction_new}}

Let $G$ be a $k$-uniform hypergraph on the vertex set $[n]$ and let $\chi = \chi(G)$.
Denoting $q = q(\kappa_{\chi,k})$, it follows that for an alphabet $F$ of size $q$ there exist an encoding function
$C: F^k \rightarrow F^\chi$ and a decoding function $D: (F \cup \{\perp\})^\chi \rightarrow F^k$ such that for every $k$-subset $e$ of $[\chi]$ and every message $m \in F^k$ it holds that $D(C_e(m)) = m$.

To prove that $q(G) \leq q$ we define a coding scheme over the alphabet $F$ as follows.
Fix a valid coloring $g:[n] \rightarrow [\chi]$ of $G$.
Consider the encoding function $\widetilde{C}:F^k \rightarrow F^n$ that given a message $m \in F^k$ outputs the vector in $F^n$ whose $i$th entry $\widetilde{C}_i(m)$ is $C_{g(i)}(m)$, i.e., the symbol in the codeword $C(m)$ which corresponds to the color of the $i$th vertex. Here, and throughout, we use the notation $C_i(m)$ to denote the $i$th entry in the codeword $C(m)$.
It remains to show that given a word $\widetilde{C}_e(m) \in (F \cup \{\perp\})^n$ for an edge $e$ of $G$ and a message $m \in F^k$, it is possible to retrieve $m$.
Indeed, consider the word $\widetilde{C}_e(m)$ restricted to the entries corresponding to an edge $e$ (recall that the value of these entries is not the erasure symbol $\perp$).
As the vertices of $e$ are colored by $g$ using distinct colors, the values of the corresponding entries in $\widetilde{C}_e(m)$ equal to $k$ distinct entries in $C(m)$. As $C(m)$ is decodable from any set of $k$ distinct entries via $D$, we conclude that given $\widetilde{C}_e(m)$ we can retrieve $m$ as well. An identical proof shows that $q_{lin}(G) \leq q_{lin}(\kappa_{\chi(G),k}) \leq [\chi(G)-1]_{pp}$ (where the rightmost inequality follows from the known upper bounds on $q_{lin}$ as discussed after Conjecture~\ref{con:MDS_linear_new}).


\subsection{Proof of Proposition~\ref{prop:Fano}}
The proof is based on the Fano plane illustrated in Figure~\ref{fig:Fano_new}.
Recall that the Fano plane is defined over a set of $7$ points, denoted here by the integers in $[7]$, and consists of $7$ lines with $3$ points on every line and $3$ lines on every point. Let $G$ be the $3$-uniform hypergraph on the vertex set $[7]$ whose edges are all the $3$-subsets of $[7]$ that do {\em not} form lines in the Fano plane. We claim that $G$ satisfies the assertion of the proposition. To this end, we turn to show that
\begin{enumerate}
  \item\label{itm:Fano1} $\chi(G) = 7$,
  \item\label{itm:Fano2} $q_{lin}(G) = q(G) = 2$, and
  \item\label{itm:Fano3} $q(\kappa_{7,3}) \geq 5$.
\end{enumerate}
For Item~\ref{itm:Fano1}, observe that every two vertices of $G$ are included in some edge of $G$, hence every valid coloring of $G$ assigns every vertex to a distinct color, implying that $\chi(G) = 7$.
For Item~\ref{itm:Fano2}, consider the three vectors $v_1 = (0,0,0,1,1,1,1)$, $v_2 = (0,1,1,1,0,0,1)$, and $v_3 = (1,1,0,1,1,0,0)$ over the binary field $F_2$, and let $C:F_2^3 \rightarrow F_2^7$ be a linear encoding function whose image is the linear span of $\{ v_1, v_2, v_3\}$.
Namely, $C(m_1,m_2,m_3)=\sum_{i=1}^3m_iv_i$.
It is straightforward to verify that if we restrict the images of the function $C$ to the coordinates of any edge of $G$ we get an invertible function from $F_2^3$ to $F_2^3$. For example, restricting $C$ to coordinates $\{1,2,4\}$ gives a function whose image is spanned by the restrictions of $\{v_1,v_2,v_3\}$ to these coordinates, i.e., a function with image spanned by $\{ (0,0,1), (0,1,1), (1,1,1)\}$. As the latter  vectors are linearly independent over $F_2$ it follows that the restriction of $C$ at hand is invertible.
We conclude that for every edge $e$ of $G$ and for every message $m \in F_2^3$, one can decode $m$ from the codeword $C(m)$ even if the symbols in the locations of $[7] \setminus e$ are erased, hence $q_{lin}(G) \leq 2$.
Since it clearly holds that $q_{lin}(G) \geq q(G) \geq 2$, Item~\ref{itm:Fano2} follows.
Finally, Item~\ref{itm:Fano3} follows from the fact that $q(\kappa_{n,k}) \geq n-k+1$ for every $n \geq k \geq 2$ (see~\cite{S60}).

\subsection{Proof of Proposition~\ref{prop:GL_new}}
Let $F$ be a field of size $q$.
Consider a $k$-uniform hypergraph $G$ whose vertices correspond to {\em normalized} vectors in $F^k$. Here, a normalized vector $(x_1,\dots,x_k) \in F^k$  is one in which the leading nonzero entry equals 1.
The number of vertices $n$ in $G$ is thus $\sum_{i=1}^k{q^{k-i}}=(q^k-1)/(q-1)$.
The edge set of $G$ consists of all $k$-collections of vertices in $G$ that correspond to linearly independent vectors.
As any two vertices in $G$ have corresponding vectors that can be completed to a linearly independent set of size $k$, any two vertices in $G$ are included in at least one edge in $G$. Thus $\chi(G)=n=(q^k-1)/(q-1)$.
In what follows, we use a natural relation between vertices in $G$ and linear functions $F^k \rightarrow F$. Specifically, if vertex $i$ in $G$ is defined by the vector $(x_1,\dots,x_k)$, then the function $f_i$ corresponding to $i$ maps $m=(m_1,\dots,m_k) \in F^k$ to $f_i(m)=\sum_{\ell=1}^{k}x_\ell m_\ell$.
Now, let $C:F^k \rightarrow F^{n}$ be a linear code which maps messages $m=(m_1,\dots,m_k)$ to the codeword $C(m)$ using the relation above. Namely, the $i$th entry $C_i(m)$ of $C(m)$ is defined to be $f_i(m)$. As edges in $G$ consist of vertices that correspond to linearly independent vectors, it follows that for every edge in $G$, the message $m$ can be recovered from $C_e(m)$. This implies that $q_{lin}(G) \leq q$.


\subsection{Proof of Lemma~\ref{lem:subgraph_new}}

For an integer $q$, let $n_q$ denote the number of vertices in the graph $G_q$.
Recall that the vertices of $G_q$ are the balanced vectors of $[q]^{q^2}$, and note that each of them can be realized as a function from $F^2$ to $F$, where $F$ is an alphabet of size $q$.
Namely, with each vertex $v$ in $G_q$ we associate a function $f_v(m)$ that takes $m \in F^2$ and returns an element in $F$. We turn to show a coding scheme over the alphabet $F$ with respect to the erasure patterns defined by the graph $G_q$.
To this end, consider the encoding function $C:F^2 \rightarrow F^{n_q}$ for which for any $m$,
$C_v(m)=f_v(m)$.
%
Here, for any vertex $v$ in $G_q$, $C_v(m)$ denotes the entry of $C$ corresponding to $v$.
For the decoding, consider a word $C(m)$ and assume that all of its symbols but the two that correspond to some adjacent vertices $u$ and $v$ are erased. We claim that given $C_u(m)$ and $C_v(m)$ it is possible to retrieve $m$.
Indeed, by the definition of $G_q$, the possible pairs $(C_u(m),C_v(m))$ over all messages $m \in F^2$ are all the distinct pairs in $F^2$, hence the pair $(C_u(m),C_v(m))$ fully determines the message $m$. This implies that $q(G_q) \leq q$.

For the second item, let $G=(V,E)$ be a graph with $q(G) = q$.
Then there exists a coding scheme over an alphabet $F$ of size $q$ with respect to the erasure patterns defined by $G$.
Let $C: F^2 \rightarrow F^{|V|}$ be the encoding function of such a coding scheme.
Observe that the existence of a corresponding decoding function implies that for every adjacent vertices $u$ and $v$ in $G$ it holds that the possible pairs $(C_u(m),C_v(m))$ over all messages $m \in F^2$ are all the pairs in $F^2$.
We assign to every non-isolated vertex $v \in V$ the vertex of $G_q$ that represents the function that assigns every $m \in F^2$ to $C_v(m)$. We further assign isolated vertices of $G$ to arbitrary vertices of $G_q$.
This mapping forms a homomorphism from $G$ to $G_q$, so in particular, it holds that $\chi(G) \leq \chi(G_q)$.

\subsection{Proof of Proposition~\ref{prop:linear:2_new}}

For an integer $q$, let $n_q$ denote the number of vertices in the graph $G_q$.
Assume in contradiction that for some prime power $q'$, $q_{lin}(G_q) = q' < [\chi(G_q)-1]_{pp}$, and let $C$ be a linear encoding function for the graph $G_q$ over a field $F$ of size $q'$.
Note that for every message $m \in F^2$, $C(m)=(C_1(m),\dots,C_{n_q}(m))$, where each $C_i(m)$ is a linear function of $m$.
One may represent each such function $C_i$ by a vector $(x_1,x_2) \in F^2$ such that $C_i(m)=x_1m_1+x_2m_2$ for every $m=(m_1,m_2)$.
By the decodability of $C$, it follows that the vectors associated with the endpoints of any edge in $G_q$ are linearly independent (in particular, there are no entries in $C$ which correspond to the zero linear function).
Thus, similar to the proof of Proposition~\ref{prop:GL_new}, we may assume that the vectors corresponding to entries of $C$ are normalized, i.e., their leading nonzero coefficient is $1$, since such a normalized $C$ is still an erasure code for $G_q$.
A simple counting argument shows that the number of distinct normalized vectors in $F^2$ is precisely $q'+1$. This implies that $\chi(G_q) \leq q'+1$, since using the distinct normalized vectors of $C$ to represent color classes one could color $G_q$. This in turn implies that $[\chi(G_q)-1]_{qq} \leq q'$, in contradiction to our assumption.

\subsection{Proposition~\ref{prop:2:nonlinear_new} and Canonical Independent Sets of $G_q$}

For an integer $q$ consider the graph $G_q$ given in Definition~\ref{def:Gq}.
For two distinct indices $i,j \in [q^2]$ let $A_{i,j}$ be the set of all vertices $u \in [q]^{q^2}$ of $G_q$ that satisfy $u_i = u_j$. Every set $A_{i,j}$ forms an independent set in $G_q$ since for every two distinct vertices $u,v \in A_{i,j}$ we have $u_i = u_j$ and $v_i = v_j$, and thus $(u_i,v_i) = (u_j,v_j)$, which implies that the collection of pairs $\{(u_i,v_i)\}_{i \in [q^2]}$ is not equal to $[q]\times [q]$. We refer to such independent sets of $G_q$ as {\em canonical}. The canonical indepdnent sets of $G_q$ are used in the proof of Proposition~\ref{prop:2:nonlinear_new} given below.

\vspace{2mm}

\begin{prf}[of Proposition~\ref{prop:2:nonlinear_new}]
Let $q$ be an integer.
By the pigeonhole principal every vector $u$ in $[q]^{q^2}$ satisfies $u_i = u_j$ for some $i<j \in [q+1]$.
This implies that the $q+1 \choose 2$ canonical independent sets $A_{i,j}$ of $G_q$ with $i<j \in [q+1]$ cover the entire vertex set of $G_q$, implying the required upper bound on its chromatic number.

For the lower bound, it suffices to show that $G_q$ contains a clique of size $n=n(q,2)$.
By definition, there exists an MDS code $C:F^2 \rightarrow F^n$ over an alphabet $F$ of size $q$.
For every $i \in [n]$, let $C_i(m)$ be the $i$th entry of $C(m)$ and consider the function $h_i:F^2 \rightarrow F$ defined by $h_i(m) = C_i(m)$.
The MDS property guarantees that for every $i \neq j \in [n]$ and every $m_1 \neq m_2 \in F^2$,
\[(h_i(m_1),h_j(m_1)) \neq (h_i(m_2),h_j(m_2)).\]
Hence, the vectors in $[q]^{q^2}$ that represent the functions $h_1,\ldots,h_n$ (with respect to an arbitrary order of $F^2$) form a clique of size $n$ in $G_q$, as desired.
\end{prf}

\BP
\label{prop:canonical}
For every sufficiently large integer $q$, the number of canonical independent sets required to cover the vertex set of $G_q$ is $\Omega(q^2)$.
\EP

\begin{prf}
Let $q$ be a sufficiently large integer. To simplify the presentation, we omit throughout the proof all floor and ceiling signs.
Let $S$ be a collection of pairs $(i,j) \in [q^2]^2$ with $i<j$ such that $|S| \leq c \cdot q^2$, where $c$ is some fixed small positive constant (say, $c < 1/4$).
Let $I \subseteq [q^2]$ be a set of size $|I|=2c \cdot q^2$ that include all the indices that appear in the pairs of $S$.
We turn to show that the union $\cup_{(i,j) \in S}{A_{i,j}}$ does not cover the entire vertex set of $G_q$.

We use the probabilistic method to construct an uncovered vertex $u$ of $G_q$ as follows.
We first pick an assignment for the entries of $u$ that correspond to the indices of $I$.
These entries are chosen uniformly at random from $[q/2]$ in a balanced manner, that is, every element of $[q/2]$ appears exactly $|I|/(q/2) = 4c \cdot q$ times.
In general, for integers $n$ and $k$ and for $i \neq j \in [n \cdot k]$, a random balanced vector $y \in [k]^{n \cdot k}$ satisfies $y_i = y_j$ with probability \[ \frac{k \cdot {nk-2 \choose n-2} \cdot {nk-n \choose n,\ldots,n}}{{nk \choose n,\ldots,n}} = \frac{n-1}{nk-1}.\]
For our case, set $k=q/2$ and $n=4c \cdot q$ to obtain that the expected number of pairs $(i,j) \in S$ such that $u_i = u_j$ is
\[ |S| \cdot \frac{n-1}{nk-1} \leq \frac{q}{2},\]
where the inequality holds assuming that $q$ is sufficiently large.
In particular, there exists a choice for the restriction of $u$ to the indices of $I$ such that for at most $q/2$ of the pairs $(i,j) \in S$ it holds that $u_i=u_j$. This allows us, for every pair $(i,j) \in S$ with $u_i=u_j$, to change the value of $u_j$ to be some distinct element from $\{\frac{q}{2}+1,\ldots,q\}$, so that the modified $u$ satisfies $u_i \neq u_j$ for every pair $(i,j) \in S$. Finally, we extend $u$ to the entries that correspond to the indices of $[q^2] \setminus I$ to obtain a balanced vector in $[q]^{q^2}$. This is possible because in our current assignment no element of $[q]$ is used more than $q$ times. We obtain a vertex $u$ of $G_q$ that does not belong to any of the independent sets $A_{i,j}$ with $(i,j) \in S$, as required.
\end{prf}

\subsection{Subgraphs of $G_q$}

\begin{definition}[The graphs $H_{q}$ and $H^{cyclic}_{q}$]\label{def:cyclic}
Let $q$ be an integer.
Let the vector representation of a permutation $\sigma \in S_q$ be $(\sigma(1),\dots,\sigma(q))$.
A vector $(v_1,\dots,v_{q^2})$ of $[q]^{q^2}$ is said to be a concatenation of $q$ permutations if for every $i \in [q]$ the vector $(v_{q(i-1)+1},\dots,v_{qi})$ is a permutation of $[q]$.
Let $H_q$ be the subgraph of $G_q$ induced by the vectors in $[q]^{q^2}$ that are concatenations of $q$ permutations.
Let $H^{cyclic}_{q}$ be the subgraph of $G_q$ induced by the vectors in $[q]^{q^2}$ that are concatenations of $q$ cyclic permutations.
\end{definition}

\BP
\label{prop:subgraphs_G_q}
For every integer $q$,
$\chi(H^{cyclic}_q) \le  \chi(H_q)\le q$.
\EP

\begin{prf}
For every $i \in [q]$, let $A_{1,q+i}$ be the set of vertices $u$ in $H_q$ such that $u_1 = u_{q+i}$. These $q$ sets cover the entire vertex set of $H_q$ and every $A_{1,q+i}$ is a canonical  independent set, implying that $\chi(H_q)\le q$. Since $H_q^{cyclic}$ is a subgraph of $H_q$ the proposition follows.
\end{prf}


\section{The $q_\eps$ Parameter}
\label{sec:q_eps}


\subsection{The study of $\kappa$.}

\BP
\label{p:q3}
$\ $

\begin{itemize}
\item For $\eps = \frac{1}{3}$, $q_{\eps}(\kappa_{20,2}) \leq 3$ whereas $q(\kappa_{20,2}) =19$, and \vspace{1mm}
\item For $\eps = \frac{1}{4}$, $q_{\eps}(\kappa_{7,2}) \leq 4$ whereas $q(\kappa_{7,2}) =6$, and \vspace{1mm}
\item For $\eps = \frac{1}{6}$, $q_{\eps}(\kappa_{6,2}) \leq 6$ whereas $q(\kappa_{6,2}) = 7$.
\end{itemize}

\EP
\begin{prf}
The values of the $q$ parameter specified in the cases of the proposition are given in \cite{CD07,Huntemann_thesis}.
Below, we construct explicit codes $C$ for the error-tolerant settings at hand.
The codes were verified by a computer program. 
Notice that if $\eps=1/q$, then for any two nodes $u,v$ in the clique at hand, the set $\{(C_u(m),C_v(m)) \mid m \in [q]^2\}$ is of size at least $q^2(1-\eps)=q^2-q$.
Otherwise, any decoding scheme will have probability of error greater than $\eps$ (over uniformly chosen messages). Here, for a node $u$, $C_u(m)$ denotes the entry of $C(m)$ corresponding to $u$.
For $q=3$ and $\kappa_{20,2}$ consider the code $C$ in which for $v=1$ to $20$ the value of $C_v(m)$ as a function of $m$ is given below in vector form.
\\ $(0,0,0,1,1,1,2,2,2)$, $(0,1,2,0,1,2,0,1,2)$, \\ $(0,0,1,0,1,2,1,2,2)$, $(0,0,1,1,0,2,2,1,2)$,\\ $(0,0,1,1,2,0,2,2,1)$, $(0,0,1,2,2,1,0,1,2)$,\\ $(0,1,0,1,2,0,2,1,2)$, $(0,1,0,2,2,1,2,0,1)$,\\$(0,1,1,0,2,2,2,1,0)$, $(0,1,1,2,0,2,0,2,1)$,\\ $(0,1,1,2,1,0,2,0,2)$, $(0,1,1,2,2,0,1,2,0)$,\\$(0,1,2,0,1,0,2,2,1)$,
 $(0,1,2,0,2,1,1,0,2)$, \\$(0,1,2,1,0,0,1,2,2)$, $(0,1,2,1,0,2,2,0,1)$,\\ $(0,1,2,1,2,1,0,2,0)$, $(0,1,2,2,0,1,2,1,0)$,\\$(0,1,2,2,1,2,1,0,0)$,
$(0,1,2,2,2,0,0,1,1)$.

Similarly for the case of $q=4$ and $\kappa_{7,2}$:\\
$(0,0,0,0,1,1,1,1,2,2,2,2,3,3,3,3)$, $(0,1,2,3,0,1,2,3,0,1,2,3,0,1,2,3)$, \\
$(0,1,2,3,1,2,3,0,2,3,0,1,3,0,1,2)$, $(0,1,2,3,2,3,0,1,1,2,3,0,3,0,1,2)$, \\
$(0,1,2,3,3,0,1,2,1,2,3,0,2,3,0,1)$, $(0,1,2,3,0,1,2,3,3,0,1,2,2,3,0,1)$, \\
$(0,1,2,3,2,3,0,1,2,3,0,1,1,2,3,0)$.

Similarly for the case of $q=6$ and $\kappa_{6,2}$:\\
$(0,0,0,0,0,0,1,1,1,1,1,1,2,2,2,2,2,2,$\\$3,3,3,3,3,3,4,4,4,4,4,4,5,5,5,5,5,5)$, \\
$(0,1,2,3,4,5,0,1,2,3,4,5,0,1,2,3,4,5,$\\$0,1,2,3,4,5,0,1,2,3,4,5,0,1,2,3,4,5)$, \\
$(0,1,2,3,4,5,1,2,3,4,5,0,2,3,4,5,0,1,$\\$3,4,5,0,1,2,4,5,0,1,2,3,5,0,1,2,3,4)$, \\
$(0,1,2,3,4,5,2,3,4,5,0,1,4,5,0,1,2,3,1$\\$,2,3,4,5,0,3,4,5,0,1,2,5,0,1,2,3,4)$, \\
$(0,1,2,3,4,5,3,4,5,0,1,2,1,2,3,4,5,0,5$\\$,0,1,2,3,4,2,3,4,5,0,1,0,1,2,3,4,5)$,  \\
$(0,1,2,3,4,5,4,5,0,1,2,3,3,4,5,0,1,2,2$\\$,3,4,5,0,1,1,2,3,4,5,0,1,2,3,4,5,0)$.
\end{prf}

\subsection{The study of general graphs $G$ ($k=2$).}

\begin{definition}[The graphs $G_{q,\eps}$, $H_{q,\eps}$, and $H^{cyclic}_{q,\eps}$]
\label{def:Gqe}

For an integer $q$ and an error parameter $\eps$, let $G_{q,\eps}$ be the graph whose vertex set consists of all the vectors of length $q^2$ over $[q]$, that is, the vectors $u \in [q]^{q^2}$, 
where two vertices $u=(u_1,\dots,u_{q^2})$ and $v=(v_1,\dots,v_{q^2})$ are adjacent if the collection of pairs $\{(u_i,v_i)\}_{i \in [q^2]}$ is at least of size $(1-\eps)q^2$.
$H_{q,\eps}$ and $H^{cyclic}_{q,\eps}$ are subgraphs of $G_{q,\eps}$ induced on the vertices of $H_q$ and $H^{cyclic}_{q}$ respectively.

\end{definition}

The graph family $G_{q,\eps}$ is extremal with respect to the $q_\eps$ parameter in the sense given by the following lemma.

\BL
\label{lem:subgraph_new_eps}
For every integer $q$,
\begin{enumerate}
  \item $q_\eps(G_{q,\eps}) \leq q$ and
  \item $\chi(G) \leq \chi(G_{q,\eps})$ for every graph $G$ with $q_\eps(G)=q$.
\end{enumerate}
\EL

\begin{prf}
For an integer $q$, let $n_q$ denote the number of vertices in the graph $G_{q,\eps}$.
Recall that the vertices of $G_{q,\eps}$ are the vectors of $[q]^{q^2}$, and note that each of them can be realized as a function from $F^2$ to $F$, where $F$ is an alphabet of size $q$.
As in the proof of Lemma~\ref{lem:subgraph_new}, we turn to show a coding scheme over the alphabet $F$ with respect to the erasure patterns defined by the graph $G_{q,\eps}$.
To this end, consider the encoding function $C:F^2 \rightarrow F^{n_q}$ that given a message $m \in F^2$ outputs the vector $C(m) \in F^{n_q}$ that consists of the evaluations of the functions that correspond to the $n_q$ vertices of $G_{q,\eps}$ on the input $m$.
For the decoding, consider a word $C(m)$ and assume that all of its symbols but the two that correspond to some adjacent vertices $u$ and $v$ are erased. We claim that given these symbols, $C_u(m)$ and $C_v(m)$, it is possible to retrieve $m$ with probability of at least $1-\eps$.
Indeed, by the definition of $G_{q,\eps}$, the possible pairs $(C_u(m),C_v(m))$ over all messages $m \in F^2$ form $(1-\eps)$ of the distinct pairs in $F^2$, hence the pair $(C_u(m),C_v(m))$ determines the message $m$ with probability of at least $1-\eps$. This implies that $q(G_q) \leq q$.

For the second item, let $G=(V,E)$ be a graph with $q_\eps(G) = q$.
Then there exists a coding scheme over an alphabet $F$ of size $q$ with respect to the erasure patterns defined by $G$.
Let $C: F^2 \rightarrow F^{|V|}$ be the encoding function of such a coding scheme.
Observe that the existence of a corresponding decoding function implies that for every adjacent vertices $u$ and $v$ in $G$ it holds that  $|\{(C_u(m),C_v(m))\mid m \in F^2\}| \geq (1-\eps)q^2$.
We assign to every non-isolated vertex $v \in V$ the vertex of $G_{q,\eps}$ that represents the function that assigns every $m \in F^2$ to $C_v(m)$. We further assign isolated vertices of $G$ to arbitrary vertices of $G_{q,\eps}$.
This mapping forms a homomorphism from $G$ to $G_{q,\eps}$, so in particular, it holds that $\chi(G) \leq \chi(G_{q,\eps})$.
\end{prf}

As with the graphs $G_q$, by Lemma~\ref{lem:subgraph_new_eps}, the challenge of obtaining graphs with chromatic number much larger than the $q_\eps$ parameter reduces to the study of the chromatic number $\chi$ of $G_{q,\eps}$ and its subgraphs.

\BP
\label{p:Gqe}
For any $q$ and $\eps=\frac{1}{q}$, $$\chi(G_{q,\eps}) \leq O\left(5^qq^{2q}\right)$$
\EP

\begin{prf}
First notice that any two vertices $u,v\in G_{q,\eps}$ for which there are $q+1$ pairs $i,j$ such that $u_i=u_j$, $v_i=v_j$ are not adjacent. By the pigeonhole principal every vertex $u$ satisfies $u_{i_1}=u_{j_1}$ for some  $i_1,j_1\in [q+1]$.
Applying the pigeonhole principal on $[q+3]\setminus\{i_1,j_1\}$, every vertex $u$ also satisfies $u_{i_2}=u_{j_2}$  for some $i_2<j_2\in [q+3]\setminus\{i_1,j_1\}$. Repeating this argument implies that every vertex $u$ satisfies  $u_{i_1}=u_{j_1},\ldots,u_{i_{q+1}}=u_{j_{q+1}}$ for some $2(q+1)$ distinct indices $i_1,j_1,\ldots,i_{q+1},j_{q+1}\in[3q+1]$. For every subset $I=\{i_1,j_1,\ldots,i_{q+1},j_{q+1}\}$ of $2(q+1)$ indices in $[3q+1]$, define the canonical independent set $$A_I=\{u\in [q]^{q^2}|u_{i_1}=u_{j_1},\ldots,u_{i_{q+1}}=u_{j_{q+1}}\}$$ of $G_{q,\eps}$. These ${{3q+1}\choose {2(q+1)}}\cdot \frac{(2(q+1))!}{2^{q+1}(q+1)!}$ independent subsets cover the entire vertex set of $G_{q,\eps}$, implying that $\chi(G_{q,\eps})\le {3q+1\choose 2(q+1)}\cdot \frac{(2(q+1))!}{2^{q+1}(q+1)!} = O(5^qq^{2q})$ 
\end{prf}

\BP
\label{p:Hqe_cyclic}
For any $q$ and $\eps=\frac{1}{q}$, $$\chi(H^{cyclic}_{q,\eps}) \le q^2.$$
\EP

\begin{prf}
We start by noticing that each vertex $v$ in $H_{q,\eps}^{cyclic}$ can be represented by an element in $[q]^q$ specifying which cyclic permutations correspond to $v$. With this representation, it is not hard to verify that two vertices $u$ and $v$ are adjacent if and only if $u-v$ has at least $q-1$ values. For every $i,j\in [q]$ let $A_{i,j}$ be the set of vertices $u \in [q]^q$ such that $u_1-u_2 = i$ and $u_1-u_3=j$. These $q^2$ sets clearly cover the vertex set, and notice that every $A_{i,j}$ is an independent set. Indeed, if $u$ and $v$ lie in $A_{i,j}$ then $u_1 – u_2 = v_1 – v_2$ and $u_1 – u_3 = v_1 – v_3$ implying that $u_1 – v_1 = u_2 – v_2$ and $u_1 – v_1 = u_3 – v_3$, so $u-v$ has at most $q-2$ values. This implies that $\chi(H_{q,\eps}^{cyclic})\le q^2$.
\end{prf}

Notice that the codes presented in the proof of Proposition~\ref{p:q3} for $q=4$ and $q=6$ (with the exception of the first codeword) correspond to subgraphs of $H_{q,1/q}^{cyclic}$.

\BP
\label{p:Hqe}
For any $q$ and $\eps=\frac{1}{q}$, $$\chi(H_{q,\eps})\le q!\cdot q.$$
\EP

\begin{prf}
For any collection of $q+1$ values $I=\{i_1,\ldots,i_{q+1}\}$ for which $i_1,i_2,\ldots,i_q$ is a permutation of $[q]$ and $i_{q+1}\in[q]$ define the set  $A_{I}=\{u\in [q]^{q^2}|u_1=u_{q+i_1},u_2=u_{q+i_2},\ldots,u_q=u_{q+i_q},u_1=u_{2q+i_{q+1}}\}$. Similar to Proposition~\ref{p:Hqe_cyclic}, each $A_I$ is an independent set and these $q!\cdot q$ independent sets cover the entire vertex set of $H_{q,\eps}$, implying that $\chi(H_{q,\eps})\le q!\cdot q$.
\end{prf}

\subsection{Average error $\epsilon$}
\label{sec:average}
In this subsection we allow average error $\epsilon$ over both the messages and the edges in $E$. Here, the decoding error is computed assuming a uniform set of $k$ messages and a uniform decoding edge in $E$.

\begin{definition}[The $q_{\tilde\eps}$ parameter]\label{def:q_avr_eps_parameter}
Let $G=([n],E)$ be a $k$-uniform hypergraph on the vertex set $[n]$ and let $\tilde\eps >0$.
Let $q_{\tilde\eps}(G)$ denote the smallest size $q$ of an alphabet $F$ for which there exist an encoding function
$C: F^k \rightarrow F^n$
and a decoding function
$D: (F \cup \{\perp\})^n \rightarrow F^k$
such that
\[\Pr_{e,m}[D(C_e(m))=m] \geq 1-\tilde\eps,\]
where $m$ is uniformly chosen from $F^k$, and $e$ is uniformly chosen from $E$.
\end{definition}
\noindent

\BT\label{the:reduction_new_avr_eps}
\label{the:average}
Let $n$ be any integer.
For any prime power $p$, $q_{\tilde\eps}(\kappa_{n,2}) \leq p$ for $\tilde\eps = \frac{1}{p+1}$ .
%
\ET

\begin{prf}
Consider first the clique $\kappa_{p+1,2}$. By Theorem~\ref{the:reduction_new}, it holds that $q(\kappa_{p+1,2}) \leq p$. Let $C$ be the corresponding code over $[p]$, where for $i \in [p+1]$, $C_i$ represents the code restricted to vertex $i$ of $\kappa_{p+1,2}$. Now for any integer $\alpha$ consider the clique $\kappa_{\alpha (p+1),2}$. Consider the code  for  $\kappa_{\alpha (p+1),2}$ obtained by labeling the vertices of  $\kappa_{\alpha (p+1),2}$ by $(i,j) \in [p+1] \times [\alpha]$ and assigning $C_i$ to vertex $(i,j)$ in $\kappa_{\alpha (p+1),2}$.  It now holds that $\Pr_{m}[D(C_e(m))=m] = 1$ for $e=((i,j),(i',j'))$ with $i \ne i'$.
%
Therefore,
\begin{align*}
\Pr_{e,m}[D(C_e(m))=m]& \geq  1- \frac{{\alpha \choose 2}(p+1)}{{\alpha (p+1) \choose 2}} = 1-\frac{\alpha-1}{\alpha(p+1)-1} \\
& \geq 1-\frac{1}{p+1}.
\end{align*}
This proves the assertion for any clique $\kappa_{\alpha (p+1),2}$ of size $\alpha(p+1)$.
For any integer $n$ consider the clique $\kappa_{n,2}$. Take $\alpha=\lfloor\frac{n}{p+1}\rfloor$, and $r=n-\alpha(p+1)$. Consider the code  for  $\kappa_{n,2}$ obtained by labeling the vertices of  $\kappa_{n,2}$ by $(i,j) \in [p+1] \times [\alpha] \cup [r] \times (\alpha+1)$ and assigning $C_i$ to vertex $(i,j)$ in $\kappa_{n,2}$.  As before, it holds that $\Pr_{m}[D(C_e(m))=m] = 1$ for $e=((i,j),(i',j'))$ with $i \ne i'$.

Therefore, using basic calculations,
\begin{align*}
\Pr_{e,m}[D(C_e(m))=m]& \geq  1- \frac{{\alpha \choose 2}(p+1)+r\alpha}{{\alpha (p+1)+r \choose 2}}
\geq 1-\frac{1}{p+1}.
\end{align*}
This proves the assertion for any clique $\kappa_{n,2}$ of size $n$.
\end{prf}




\end{document}